\documentstyle[preprint,tighten,aps]{revtex} 
  \begin{document}

  \draft
  \preprint{\vbox{\noindent
  Submitted to Physics Letters B \hfill INFNCA-TH9805   \\
                           \null \hfill INFNFE-0898     \\
                           \null \hfill astro-ph/9807078}}
  %
  \title{Helioseismology can test the Maxwell-Boltzmann distribution}
  \author{S.~Degl'Innocenti$^{1,2,}$\cite{email1},
          G.~Fiorentini$^{1,3,}$\cite{email2}, 
          M.~Lissia$^{4,5,}$\cite{email3},
	  P.~Quarati$^{6,4,}$\cite{email4}, 
          B.~Ricci$^{1,3,}$\cite{email5}
         }
  \address{
  $^{1}$Istituto Nazionale di Fisica Nucleare, Sezione di Ferrara,
        via Paradiso 12, I-44100 Ferrara, Italy\\
  $^{2}$Dipartimento di Fisica dell'Universit\`a di Pisa, p.zza Torricelli 1,
        I-56100 Pisa, Italy\\
  $^{3}$Dipartimento di Fisica dell'Universit\`a di Ferrara,
        via Paradiso 12, I-44100 Ferrara, Italy\\
  $^{4}$Istituto Nazionale di Fisica Nucleare, Sezione di Cagliari,\\
        Cittadella Universitaria, I-09042 Monserrato, Italy\\
  $^{5}$Dipartimento di Fisica dell'Universit\`a di Cagliari,\\
        Cittadella Universitaria, I-09042 Monserrato, Italy\\
  $^{6}$Dipartimento di Fisica, Politecnico di Torino, I-10129 Torino, Italy
          }
  \date{8 July 1998}  
  \maketitle
  \begin {abstract}
  Nuclear reactions in stars occur between nuclei in the high-energy tail
  of the energy distribution and are sensitive to possible deviations from
  the standard equilibrium thermal-energy distribution. We are able to
  derive strong constraints on such deviations by using the detailed
  helioseismic information of the solar structure. If a small deviation
  is parameterized with a factor $\exp\{ -\delta (E/kT)^2\}$, we find that
  $\delta$ should lie between -0.005 and +0.002. However, even
  values of $\delta$ as small as $0.003$ would still give important
  effects on the neutrino fluxes.
  \end {abstract}

  \narrowtext

  \subsection{Introduction}
  \label{sec:intr}
   Thermal averages are fundamental ingredients of the theoretical
   description of many physical phenomena: solar modeling is a specific
   example.
   These thermal averages can and are often described as integrals
   weighted by the appropriate equilibrium distribution functions; {\em e.g.},
   the mean square averaged velocity $\langle v^2\rangle$ of a particle
   in a gas is obtained from the integral of $v^2$ times the one-body velocity
   distribution function $f(v)$, the pressure from the integral of $p v$
   times $f(v,p)$ ($p$ is the momentum),
   and so on~\cite{landauV59,huang63}.
  
   In the limit of non-interacting states, infinite volume and zero
   density, a single scale
   (the temperature or the average one-body energy) characterizes
   all the equilibrium distributions, which are described by the
   Maxwell-Boltzmann distribution (MBD). It is well-known that, even for
   non-interacting states, when the system is finite and/or the density is
   not zero, the distribution deviates form the MBD, and the resulting
    distribution (microcanonical, Fermi-Dirac or Bose-Einstein) is
   characterized by additional scales (total energy, Fermi energy, etc.).
   Similarly, the interaction could produce additional dynamical scales
   that modify the free distribution,
   {\em e.g.}, $^4$He is phenomenologically better described as a
   weakly-interacting Bose system than in terms of its fermionic constituents.
   In principle the thermal distribution of the effective weakly interacting
   degrees of freedom (the bosonic $^4$He nuclei, in this example) could be
   dynamically calculated from the original strong interacting elementary
   particles (the nucleons).
   However, theoretical calculations of thermal distribution
   functions for strongly interacting systems are very difficult, and
   one often resorts to phenomenologically motivated parameterizations. In
   specific cases, it has been possible to derive {\em equilibrium} 
   distributions departing from the MBD~\cite{kaniadakis93,kaniadakis97},
   and, more in general, theoretical
   frameworks~\cite{tsallis88,curado91,tsallis95c}
   have been formulated 
   that naturally produce nonstandard distributions.
  
   In spite of this, one can argue that, even in presence of strong
   and/or many-body and/or long-range forces, one single scale dominates
   in many practical case and, therefore, that the MBD
   is an excellent approximation. This argument is confirmed {\em a
   posteriori} by its phenomenological success. However, one should
   also keep in mind that many applications do not test the details of the
   distributions, but only one or a few moments.
   In particular, if a physical quantity is determined by only one moment,
   one can always summarize the relevant information in the most economical
   way by using the MBD.
   In practice, in many cases nothing changes using distributions that
   differ only in the higher moments.
   This low sensitivity of many important physical observables to
   the details of the thermal distribution together with the difficulties
   of the microscopic calculation leads to the possibility of considering
   more general distributions that depart from the MBD.
   
   For instance, already two decades 
   ago~\cite{kocha7475,clayton74,clayton75} 
   it was proposed that
   small depletions of the high-energy tail of the relative energy
   distribution could modify the solar neutrino fluxes. This same idea has
   been recently 
   reconsidered~\cite{haubold95,kania96,quarati97,kania97}
   in the light of the new developments that
   put nonstandard equilibrium distributions on a firmer ground.
  
Similarly, one could invoke a small enhancement of the high-energy tail
of the proton distribution in order to efficiently burn lithium near the
bottom of the convective zone. This could be regarded as
 an attempt to account for the low photospheric lithium abundance
 (about a factor 100 lower than the meteoric one~\cite{AG89}),
 essentially unexplained  within the Standard Solar Model (SSM)
 (see, however, Ref.~\cite{RCVD96}).

   In this paper we take the opposite approach and study what kind of
   constraints our best knowledge of solar physics, both theoretical and
   observational, can impose on possible deviations from standard
   thermal distribution.
   Our two basic tools will be the sub-barrier fusion reaction rates and
   helioseismology.

  {\em Nuclear reactions in stars occur
   generally between nuclei with kinetic energy much larger than $kT$
   and are  thus suitable for probing the high-energy tail of the particle
   distribution.}  
   Even for the $pp$ reaction, which has the lowest
   barrier, the Gamow peak in the solar core is at energy about
   five times larger than $kT$,
   making the reaction rate very sensitive to changes in number of
   high-energy particles. Therefore, if we can precisely determine a reaction
   rate by means of observations,
   this determination can be used to constrain the particle energy distribution.
   Stellar interiors are indeed an ideal laboratory for this investigation:
   they are to a very high degree in thermal equilibrium, and
   the density is high enough to make deviations from standard statistics
   conceivable.
      
   Moreover, helioseismology allows us to look deeply 
   in the core of the Sun. 
  The extremely precise measurements of a tremendous number of frequencies
   enable us to extract values of sound speed with high
   accuracy even near the solar
   center~\cite{eliosnoi,BT96,BCD97,G96,BPBCD97,DGPS94}.
   In addition several properties of the convective envelope
   are accurately determined by means of helioseismology.

   Recent standard solar models that include the state-of-the-art
   ``standard'' solar physics  are in good agreement with
   helioseismic data~\cite{eliosnoi,BPBCD97,RCVD96}.
   These solar models implicitly
   assume that the solar core can be described in terms of a gas of
   particles interacting via two-body forces with no many-body
   effects apart for mean-field screening. In particular, the ion relative
   velocity distribution follows the MBD and the diffusion of the average
   number of particles is Brownian. In some sense, helioseismology tells
   us that this framework is basically correct. Nevertheless, it is important
   to quantitatively assess to what extent nonstandard distributions are
   still compatible with present data.
   In this respect we remark that {\em the information on the solar interior 
   provided by helioseismology is so detailed that 
   the $pp$ reaction rate can be reliably constrained}~\cite{eliospp}.

We shall investigate solar models obtained by modifying a SSM
 so as to include the effects of a nonstandard
distribution. By requiring that the predictions of the
resulting solar models agree with the helioseismic determinations
of convective envelope properties, we shall constrain the possible
nonstandard distributions.

\subsection{Modified statistics and burning rates}
\label{secstat}

In the ordinary treatment, the single particle energy distribution
is taken as a MBD for protons and other
ions~\footnote{In the solar plasma quantum corrections to the
statistics are relevant for electrons but not for nuclei.}:
\begin{equation}
f_{MBD}(E) =\frac{2} {\sqrt{\pi}} \,
\, \frac{\sqrt{E}}{(kT)^{3/2}} \, e^{-E/kT} \, .
\end{equation}

The nuclear burning rate between
nuclei with mass numbers $i$ and $j$ is given by:
\begin{equation}
 \label{sigmav}
 \langle \sigma v \rangle _{ij} =
 \int d\epsilon \, \sigma_{ij}(\epsilon)  \, 
f_{ij}(\epsilon) v(\epsilon) \, ,
\end{equation}
where $\epsilon$ is the collision energy, $v$ is the
relative velocity,  the cross section $\sigma$
has the usual parameterization
\begin{equation}
\sigma_{ij}(\epsilon) = \frac{S_{ij}}{\epsilon}
 e^{- b Z_i Z_j \sqrt{\mu/\epsilon}} \, ,
\end{equation}
and $f_{ij}(\epsilon)$ is the collision-energy distribution of the reacting
nuclei.
As well known, if the one particle distribution is a MBD, so it is
the collision-energy distribution $f_{ij}(\epsilon)$. 

Small deviations of $f_{ij}(\epsilon)$ from the MBD can be parameterized to
first approximation by introducing a dimensionless parameter 
$\delta$
\begin{equation}
\label{fdelta}
f^{(\delta)}_{ij}(\epsilon)
=f_{MBD}(\epsilon) \, e^{-\delta(\epsilon/kT)^2} \, ,
\end{equation}
so that for $\delta$=0 the classical statistics is
recovered~\cite{clayton74,clayton75,kania97};
for small $\delta$ the distribution is close to the standard one at
values of $\epsilon$ near  the thermal energy $kT$,
whereas significant distortion occurs in the high-energy tail.
For $\delta > 0$ this parameterization implies a depletion of the 
tail. The same parameterization can also be used to mimic
an enhanced tail ($\delta<0$), understanding that a suitable cutoff is
introduced~\cite{kania97}.

Solar models corresponding to modified statistics have been
built by using our stellar evolutionary code FRANEC~\cite{ciacio97},
 where all the nuclear-reaction rates have been calculated
according to Eqs.~(\ref{sigmav}) and (\ref{fdelta}). 
We remark that we are assuming here that $\delta$ is the same
for every reaction and is constant in the nuclear energy production
region ($R/R_\odot \leq 0.2$). 

The solar structure is  primarily sensitive  to the rate 
of just two reactions: 

\noindent
i)  $p+p \rightarrow d +e^+ + \nu_e$. Since this reaction
is at the
 basis  of the nuclear-reaction chain that sustains
 the Sun against gravitational collapse, it is natural that the internal
 solar structure is strongly influenced by its rate.
As shown in Ref.~\cite{eliospp}, this rate is strongly constrained by 
helioseismic  determinations of the convective envelope.

\noindent
ii) $p+^{14}N \rightarrow ^{15}O + \gamma$. The rate of this reaction
governs the efficiency of the CNO cycle, which is marginal 
according to the SSM. 
An enhancement of the high energy tail ($\delta < 0$)
 makes the CNO cycle  more  efficient and even dominant,
resulting in solar models drastically different from 
the SSM ({\em e.g.}, the energy production is concentrated near the
center, a convective core can arise, \dots).

\subsection{Modified statistics and the properties of 
the convective envelope}
\label{secconv}

As shown in Ref.~\cite{eliosnoi}, helioseismology determines
with high accuracy three independent properties  $Q$ of
the convective envelope:
its depth $R_b$, the density at its bottom $\rho_b $
and the photospheric helium abundance $Y_{ph}$.

   The helioseismic values of these quantities are shown in
Table~\ref{tabhelios} together with two estimates of their uncertainties:
$(\Delta Q/Q)_{cons}$ corresponds to the (very) conservative
definition of Ref.~\cite{eliosnoi}, whereas $(\Delta Q/Q)_{1\sigma}$
is the corresponding $1\sigma$ ``statistical'' error estimate.

In the same Table we also show the predictions of 
the ``model with helium and heavy elements diffusion''
of Ref.~\cite{bahcall95a} (BP95), 
which are in excellent agreement with the
helioseismic determinations. For this reason we shall use this model as the
reference SSM. As an example of possible ``systematic'' theoretical
uncertainties, we also show results from our solar model including
helium and heavy elements diffusion~\cite{ciacio97} (FR97),
which deviates somewhat from the helioseismic determinations.

By numerical experiments with FR97, we have also determined the
dependence of these three properties on $\delta$; results are shown
in Fig.~\ref{figdelta}. The different behavior for negative and
positive $\delta$'s becomes more evident as $|\delta|$ increases.
This qualitative difference reflects a physical effect: when
the tail of the distribution is enhanced ($\delta<0$)
the  CNO cycle becomes important (at $\delta=-0.01$ 
the contributions of the pp-chain and of the CNO cycle
are about the same). Since the Gamow energy for the $p+^{14}N$ reaction
near the solar center is about 27 keV, a factor five 
larger than the one for the $pp$ reaction, even a small
$\delta$  yields  drastic effects on the reaction rate
and, consequently, on the solar structure.

Nonetheless, for small values of $\delta$, these dependences can be
parameterized by power laws:
\begin{eqnarray}
\label{leggedelta}
\frac{Q_i}{Q_i^{\text{SSM}}} &=& 
\left ( e^{-\delta} \right )^{\alpha_{Q_i}} \, ,
\end{eqnarray}
where the constant exponents $\alpha_{Q_i}$ are shown in the last column of
Table~\ref{tabhelios}. The solid curves in Fig.~\ref{figdelta}
demonstrate the goodness of such a parameterization in the range of
$\delta$ that is relevant to our results ($|\delta|<0.005$).

Our basic strategy will be the following:
{\em we determine the acceptable range of $\delta$ such that $R_b$, $\rho_b$
and $Y_{ph}$ are predicted within their helioseismic ranges}, by using
Eqs.~(\ref{leggedelta}) to determine the dependence of these properties
on $\delta$.

There are at least four major uncertainties in building standard
solar models that also have the potentiality of affecting the three
helioseismologic properties under investigation, and, therefore, that could
interfere with/hinder the effect of $\delta$: the astrophysical factor
$S_{pp}$, the solar opacity $\kappa$, the heavy element abundance
$\zeta = Z/X$, and the diffusion coefficients. We shall add all these effects
one after the other, and determine a range of $\delta$'s that
takes into account these uncertainties.

\subsection{Results}
\label{secresult}

For determining the range of $\delta$ allowed by helioseismology,
we use several approaches corresponding to increasing
conservativeness.
We start by defining a $\chi^2$ as:
\begin{equation}
\label{chi2_0}
\chi^2(\delta) = \sum_i 
\left ( \frac{Q_i(\delta)-Q_{\odot i}}{\Delta Q_i} \right )^{2} \, ,
\end{equation}
where $Q_i (\delta)$ are computed by using Eqs.~(\ref{leggedelta})
and the errors are the $1\sigma$ estimate of 
Table~\ref{tabhelios}. The value $\chi^2(0)$ indicates how well the SSM
reproduces these  helioseismic properties. The first row of
Table~\ref{tabdelta} shows the good agreement between  BP95 
and helioseismology ($\chi^2$/dof = 8.61/3). 

If we use $\delta$ as free parameter (second row of Table~\ref{tabdelta}),
we find the following best fit value ($\chi^2$/dof = 0.08/2) and $1\sigma$
range:
\begin{equation}
\label{rangedeltanospp}
  \delta = (-0.77 \pm 0.26) \times 10^{-3} \, .
\end{equation}
These strict constraints on the allowed values of $\delta$ come mainly
from the precise determination of density at the bottom of the convective
envelope combined with its strong dependence on $\delta$. 
In fact, the relative change of $Q_i$  due to $\Delta\delta$ is
approximately $\Delta\delta \times \alpha_{Q_i} $; therefore, the
allowed variations of $\delta$ can be estimated as
$\Delta\delta \sim 
( \Delta Q_i / Q_i )_{1\sigma} / |\alpha_{Q_i}|$ (last column of
Table~\ref{tabhelios}).

\subsubsection{Uncertainties on $S_{pp}$}
A conservative estimate of the uncertainty is provided by the range of
the published results~\cite{NACRE}, whereas
a $1\sigma$ estimate has been provided in~\cite{KB94}; we shall use
$\Delta S_{pp}/S_{pp}^{\text{SSM}} = 0.05/3$ at $1\sigma$
(5\% is the ``$3\sigma$ error'' estimate).
The dependence of $Q_i$ on $S_{pp}$ has been determined numerically in
Ref.~\cite{eliospp}. By redefining a suitable $\chi^2(\delta,S_{pp})$:
\begin{equation}
\label{chi2spp}
\chi^2(\delta, S_{pp}) = \sum_i
\left ( \frac{Q_i(\delta,S_{pp}) - 
                Q_{\odot i}}{\Delta Q_i} \right )^{2} \, + \,
\left ( \frac{S_{pp}- S_{pp,\text{SSM}}}{\Delta S_{pp}} \right )^{2}\, ,
\end{equation}
we find that the best fit value of $\delta$ does not change and that
the $1\sigma$ range is double:
\begin{equation}
\label{rangedeltaspp}
  \delta = (-0.77 \pm 0.50)\times 10^{-3} \, , 
\end{equation}
and that, consistently, the best fit value for $S_{pp}$ is
$S_{pp,\text{SSM}}$. The facts that the SSM is already
in very good agreement with helioseismology and that the dependence of
the $Q_i$ on $S_{pp}$ is much weaker than that on $\delta$ explain these
results.

\subsubsection{Uncertainties on $\kappa$ and $\zeta$}
The heavy element abundance $\zeta$ and the solar opacity $\kappa$
are known with a conservative accuracy of about
$ 10$\%~\cite{BP92,bahcall95a,castell97}. Therefore,
our $1\sigma$ relative error estimate will be $ 0.1/3$.
The dependence\footnote{We remark that we are considering a constant rescaling
of opacity along the solar profile.} of $Q_i$ on $\kappa$ and $\zeta$ 
has been determined numerically in Ref.~\cite{eliospp}. In this case,
the relevant $\chi^2$ is:
\begin{eqnarray}
\label{chi2sppzk}
\chi^2(\delta, S_{pp},\zeta,\kappa) &=& \sum_i
\left ( \frac{Q_i(\delta,S_{pp},\zeta,\kappa) - 
                Q_{\odot i}}{\Delta Q_i} \right )^{2} \\
 &+&
\left ( \frac{S_{pp}- S_{pp,\text{SSM}}}{\Delta S_{pp}} \right )^{2} \, + \,
\left ( \frac{\kappa-\kappa_{SSM}}{\Delta \kappa} \right )^{2} \, + \,
\left ( \frac{\zeta -\zeta_{SSM}}{\Delta \zeta} \right ) ^{2} \, .
\end{eqnarray}
We find a small change of the best fit value and, again, an
increase of the $1\sigma$ range of $\delta$:
\begin{equation}
\label{rangedeltasppzd}
  \delta = (-0.75 \pm 0.67)\times 10^{-3} \, .
\end{equation}
Comparing the 3th 4th and 5th row of Table~\ref{tabdelta}, one can notice
that most of the effect is due to $\zeta$.

\subsubsection{Uncertainties on diffusion coefficients}
We use a SSM that includes element diffusion,
calculated by solving the Burgers equation~\cite{thoule94}.
Indeed, diffusion has been
an essential ingredient of stellar evolutionary codes for achieving
agreement between predicted and helioseismic values of properties of the
convective envelope~\cite{eliosnoi}.
The success of solar models with diffusion, and the corresponding failures
of models  that neglect diffusion, suggest that the diffusion process has
been properly treated. However, in spite of the extensive discussion
about the many assumptions underlying the calculation
method~\cite{cox89,thoule94,bahcall95a}, no quantitative estimate of the
uncertainties of the calculated diffusion coefficients has been
presented.

Therefore, we also allow the diffusion efficiency to vary freely by rescaling
the diffusion coefficients by an overall constant factor $D$
($D=1$ corresponds to the SSM). We have determined the appropriate scaling
laws~\cite{eliosdiff} of the properties of the convective zone.
For completeness, we report the complete dependence on all the considered
quantities ($\delta$, $S_{pp}$, $\zeta$, $\kappa$, and $D$):
\begin{mathletters}
\begin{eqnarray}
\label{leggi2a0}
\frac{R_b}{R_{b, \text{SSM}}} &=&
\left ( e^{-\delta} \right )^{-2.2}
\left ( \frac{S_{pp}}{S_{pp}^{\text{SSM}}} \right )^{-0.058} \,
\left ( \frac{\kappa}{\kappa_{\text{SSM}}} \right )^{-0.0084} \,
\left ( \frac{\zeta}{\zeta_{\text{SSM}}} \right )^{-0.046} \, 
D^{-0.016}
\\
\label{leggi2b0}
\frac {\rho_b}{\rho_{b,\text{SSM}}} &=&
\left ( e^{-\delta} \right )^{33.6}
\left ( \frac{S_{pp}}{S_{pp}^{\text{SSM}}} \right )^{0.86} \,
\left ( \frac{\kappa}{\kappa_{\text{SSM}}} \right )^{0.095} \,
\left ( \frac{\zeta}{\zeta_{\text{SSM}}} \right ) ^{0.47} \, 
D^{0.14}
\\
\label{leggi2c0}
\frac{Y_{ph}}{Y_{ph, \text{SSM}}} &=&
\left ( e^{-\delta} \right )^{6.2} 
\left ( \frac{S_{pp}}{S_{pp}^{\text{SSM}}} \right )^{0.14} \,
\left ( \frac{\kappa}{\kappa_{\text{SSM}}} \right )^{0.61} \,
\left ( \frac{\zeta}{\zeta_{\text{SSM}}} \right ) ^{0.31} \,
D^{-0.091} \, .
\end{eqnarray}
\end{mathletters}

No additional term is added to $\chi^2$, since we assume that $D$ is
completely undetermined (infinite error), and we let it vary freely.
The only dependence of $\chi^2$ on $D$ is through 
$Q_i(\delta,S_{pp},\zeta,\kappa,D)$. As it is shown in the last row of
Table~\ref{tabdelta},  the $1\sigma$ allowed range becomes:
\begin{equation}
\label{rangedelta3sppd}
\label{rangeBP}
  \delta = (-0.91 \pm 1.06)\times 10^{-3} \, ,
\end{equation}
and we find that the best fit value for $D$ is only 3\% smaller than the
standard one.

\subsubsection{Solar model ``theoretical uncertainties''}
At last we try to estimate how much our results could depend on having used
BP95 as reference standard model. 
To this end, we consider one of the standard solar models (models that
include all the state-of-the-art solar physics), whose helioseismic properties
differ the most from BP95 and, consequently, fit less well the
experimental data.
We repeated the above-described analysis by using FR97 as
standard solar model. When all parameters are varied, the $1\sigma$ range 
and best fit value, {\em cf}. Eq.~(\ref{rangeBP}), become:
\begin{equation}
\label{rangedeltaf97}
  \delta = (-1.79 \pm 1.04)\times 10^{-3} \, .
\end{equation}
The corresponding fit to the helioseismic properties is acceptable
($\chi^2$/dof = 2.32/1), the best fit value for $D$ is 9\% smaller
than the standard one, and the values for $\zeta$ and $\kappa$ are
2\% larger.

\subsection{Discussion and conclusions}
\label{secdis}
The constraints we have found on non-Maxwellian statistics look rather
strict, as the dimensionless parameter cannot exceed a few per thousand.
In fact, if we define a conservative interval as the union of the
$3\sigma$ ranges found by using BP95 and FR97 SSMs, we find
\begin{equation}
\label{rangedelta3s}
  -4.9 \times 10^{-3} < \delta < 2.3 \times 10^{-3} \, .
\end{equation}

However, even these small values of $\delta$ could have non-negligible
implications for those observables that are sensible to the high-energy
tail of the distribution. As an example, we have estimated
the possible effects of $\delta = \pm 3\times 10^{-3}$ in the two
cases mentioned in the introduction.

\subsubsection{Neutrino fluxes}
In Table~\ref{tabflussi}, we report the effect of nonstandard statistics
on the main fluxes and on the signals of the chlorine and gallium
radiochemical experiments. Even for such small values of $\delta$ the
{\em boron and beryllium fluxes change substantially}.

\subsubsection{Lithium abundance}
As well known, the photospheric abundance of lithium is a factor about 100
lower compared to the meteoric one~\cite{AG89}. Different mechanisms have
been proposed to explain this
depletion~\cite{RCVD96,straniero96,CDP95}.
Let us discuss the possibility that nonstandard velocity distribution
could contribute to this depletion.
First of all we note that, since the lithium abundance should be
reduced in order to solve/alleviate the problem, the lithium burning rate
should be enhanced relative to the standard case. This is achieved by a
longer high-energy tail, {\em i.e.}, $\delta <0$.

We assume that the limits on $\delta$ derived in the production region
apply also up to the bottom of the convective zone, and consider
$\delta = -3\times 10^{-3}$. This value of $\delta$ yields a reduction
of the $^7$Li abundance by only 7\%, where the characteristics of the
bottom of the convective zone has been taken from FR97 
($T_b=2.1 \times 10^6$~K, $\rho_b=0.18$~g/cm$^3$ and 
$X=0.744$). Depletions comparable with the observed ones could be
obtained with $\delta \sim -0.15$, a value well outside the range
reported in Eq.~(\ref{rangedelta3s}).

\begin{table}
\caption[errori]{
The three independent properties of the convective envelope used in
our analysis. The first column ($Q_i$) labels the property, the second (BP95)
and third (FR97) columns show the values predicted by the
reference solar model BP95~\cite{bahcall95a}
and by FR97~\cite{ciacio97} standard model, the fourth column
($Q_{i\odot}$) shows the value derived by helioseismic measurements,
and the next two columns the corresponding conservative and $1\sigma$ errors.
The last two columns show the exponents that determine the dependence from
$\delta$, $\alpha_{Q_i}\equiv - d\log{Q_i}/d\delta $, and the ratios
between the $1\sigma$ error and $|\alpha_{Q_i}|$ (see text).
               }
\begin{tabular}{lccccccc}
$Q_i$ &  BP95  & FR97 & $Q_{i\odot}$ & $(\Delta Q_i / Q_i )_{\text{cons}}$
               & $( \Delta Q_i / Q_i )_{1\sigma}$ &  
         $\alpha_{Q_i}$ & $( \Delta Q_i / Q_i )_{1\sigma} / |\alpha_{Q_i}|
           \times 10^3 $\\
\hline
Y$_{ph}$     & 0.24695 & 0.2321 & 0.249 &  0.042\phantom{0}
             & 0.014\phantom{0} & 6.2 & 2.2 \\
$R_b/R_\odot$& 0.712\phantom{00} & 0.715\phantom{0} & 0.711 & 0.004\phantom{0}
             & 0.002\phantom{0} & -2.2 & 0.9\\
$\rho_b$ [g/cm$^3$] & 0.187\phantom{00} & 0.182\phantom{0} & 0.192
             & 0.037\phantom{0} & 0.0094 & 33.6 & 0.3 \\
\end{tabular}
\label{tabhelios}
\end{table}

\begin{table}
\caption[delta]{Deviations from standard statistics allowed by helioseismic
measurements. The first five columns show whether the parameter is kept
fixed (F) at its SSM value or it is allowed to vary (V) as a free parameter
within the range discussed in the text. The sixth
column shows the resulting $\chi^2$ per degree of freedom. The last two
columns show the best fit value for $\delta$ and its $1\sigma$ error. 
 }
\begin{tabular}{cccccddd}
$D$ & $\kappa$ & $\zeta$ & $S_{pp}$ & $\delta $ & $\chi^2$/dof &
           $\delta_{\text{Best}} \times 10^{3}$ & $\Delta \delta $ \\
\hline
 F  &  F  &  F  &  F  & F  &  8.61 /3  &       &      \\
 F  &  F  &  F  &  F  & V  &  0.08 /2  & -0.77 & 0.26 \\
 F  &  F  &  F  &  V  & V  &  0.08 /2  & -0.77 & 0.50 \\
 F  &  F  &  V  &  V  & V  &  0.08 /2  & -0.73 & 0.67 \\
 F  &  V  &  V  &  V  & V  &  0.04 /2  & -0.75 & 0.67 \\
 V  &  V  &  V  &  V  & V  &  0.001/1  & -0.91 & 1.06 \\
\end{tabular}
\label{tabdelta}
\end{table}

\begin{table}
\caption[ttt]{Effects of nonstandard statistics on neutrino fluxes.
Relative deviations from SSMs of the $^7$Be and $^8$B neutrino fluxes and
of the expected signals for gallium and chlorine detectors 
in two nonstandard solar models with deformed
velocity distribution ($\delta \neq 0)$.
             }
\begin{tabular}{ldd}
         & $\delta = +3\times 10^{-3}$ & $\delta = -3\times 10^{-3}$ \\
\hline
$ \Delta \Phi_{\text{Be}} / \Phi_{\text{Be}} $ & -0.30  & +0.38   \\ 
$ \Delta \Phi_{\text{B}}  / \Phi_{\text{B}}  $ & -0.55  & +1.15   \\ 
\hline
$\Delta S_{\text{Cl}} / S_{\text{Cl}} $        & -0.16  & +0.31   \\
$\Delta S_{\text{Ga}} / S_{\text{Ga}} $        & -0.50  & +1.03   \\
\end{tabular}
\label{tabflussi}
\end{table}

\begin{figure}
\caption[b]{Dependence on $\delta$ of the three independent properties of
the convective envelope used in our analysis. Crosses show the values
of the photospheric helium abundance $Y_{ph}$ (a), the density at the
bottom of the convective envelope $\rho_{b}$ (b), and the depth of this
envelope $R_{b}$ (c), relative to their standard values, as functions
of $\delta$. The solid curves are the fits in Eq.~(\ref{leggedelta})
with the exponents $\alpha_{Q_i}$ from the second last column of
Table~\ref{tabhelios}.
           }
\label{figdelta}
\end{figure}

\end{document}